%% file: pcaselli.tex
\documentclass[multphys,vecphys]{svmult}

\usepackage{makeidx}         
\usepackage{graphicx}        

\makeindex             

\begin{document}

\title*{Chemical processes in star forming regions}
\author{Paola Caselli}
\institute{INAF--Osservatorio Astrofisico di Arcetri, Largo E. Fermi 5,
I--50125 Firenze, Italy
\texttt{caselli@arcetri.astro.it}}
%
%
\maketitle

This paper will review the basic concepts of gas--phase and grain surface
chemistry of dense molecular clouds, where low mass and high mass 
stars form.  The chemistry
 of cold pre-stellar cloud cores, where molecular freeze-out
and deuterium fractionation dominate, will be presented.  Then,
following cloud evolution after protostellar birth, hot core and shock
chemistry will be discussed in view of recent observations. A brief
summary of the chemistry in protoplanetary disks will also be
furnished.  The aim is to identify important gas tracers in the
various steps of the star formation process, pointing out the main 
problems still open in the field of astrochemistry.

\keywordname{astrochemistry; Stars: formation; ISM: clouds; ISM: molecules}

\section{Introduction}
\label{sec:intro}

A good comprehension of the physical 
properties of molecular clouds and star forming regions implies
a detailed radiative transfer analysis of molecular line data 
and their interpretation with chemical models, inclusive of gas--phase reactions
as well as gas--grain and grain surface processes.  In fact, 
molecular lines allow us to determine volume densities and 
temperatures (e.g. Evans 1999), independently from dust continuum emission and 
absorption measurements (e.g. Ward--Thompson et al. 1999; Bacmann et al. 2000;
Visser et al. 2002). 
This is important to refine the results derived
from the dust continuum observations, which are affected by (factor of 
a few) uncertainties due to the poorly known properties of dust grains (e.g. 
coagulation degree, thickness of icy mantles; Ossenkopf \& Henning 1994; see
also Bianchi et al. 2003) and the presence of dust temperature gradients 
(e.g. Evans et al. 2001; Galli et al. 2002; Pagani et al. 2004). 
Furthermore, molecular lines offer a unique way to study the 
kinematics of dense cores (e.g. Ohashi et al. 
1999; Lee et al. 2001; Shinnaga et al. 2004). 

The number of known interstellar molecules is continuously growing
(see http://astrochemistry.net for a constantly updated 
list of interstellar molecules).    
Up to now, 137 molecules have been detected in the 
interstellar medium (205 including isotopomers) and about 50 in comets. 
The largest species so far observed contains 13 atoms 
(HC$_{11}$N; Bell \& Matthews 1985). 
The present paper will not be an exhaustive review of the whole
field of astrochemistry. It has the aim of introducing the basic
concepts of gas phase and surface chemistry (Sect.~\ref{sec:basic}) 
and showing what are the main chemical processes in star forming regions
(Sect.~\ref{sec:star}),
from dark clouds to protoplanetary disks. The hope is to highlight
the beauty of this field, which has become a crucial ingredient 
in all the areas of astrophysics.  Interested readers are invited 
to see the following reviews: van Dishoeck \& 
Blake (1998), Evans (1999), Ehrenfreund \& Charnley (2000), van Dishoeck (2004). 
Useful books are ``The chemically controlled cosmos'' by Hartquist 
\& Williams (1995); ``Interstellar Chemistry'' by Duley \& Williams (1984); 
``The formation of stars'' by Stahler \& Palla (2004); ``The Physics and
Chemistry of the Interstellar Medium'' by Tielens (2005). 

\section{Basic concepts in astrochemistry}
\label{sec:basic}

The ingredients of astrochemical models are: atoms (with
elemental abundances with respect to hydrogen nuclei  
reported in Tab.~\ref{tab:element}), molecular species, electrons, 
dust grains, cosmic rays and UV photons. Models of dark clouds typically 
neglect the action of UV photons (which is a good assumption when 
the visual extinction is larger than $\sim$ 4 mag; e.g. McKee 1989),
and start with all the elements in atomic form (although sometimes 
hydrogen is assumed to be in molecular form initially; e.g. Leung et al. 1984).

The first problem that chemists have to face concerns the elemental abundances.  
Four different values are listed in Tab.~\ref{tab:element} 
for each atomic species, two measured and two typically adopted in chemical 
models: (i) the {\it solar} abundance, 
measured in the solar photosphere (from Anders \& Grevesse 1989, but 
see Asplund et al. 2005 
for recent revisions on these values); (ii) the {\it cosmic} abundance, 
slightly but significantly (factor between 1.5 and 2) lower than the solar 
values, measured in the direction of stars in the solar neighborhood 
(Snow \& Witt 1996); (iii) the {\it high metal} abundances, used for 
diffuse cloud models, in which elements 
are depleted as found in diffuse clouds according to Morton (1975); 
(iv) the {\it low metal} abundances, used 
for dense cloud chemistry,
where the elements S, Si, Na, Mg, and Fe are further depleted by two 
orders of magnitude.  The {\it low metal} mix has been used to best 
reproduce the observed abundances of sulfur--containing species 
(Prasad \& Huntress 1982), complex molecules (Herbst \& Leung 1989), 
and the ionization balance of interstellar clouds
(Graedel et al. 1982).  

A better way to proceed
would be to start with diffuse--cloud conditions
and follow elemental depletion with time until dark cloud conditions
are reached; however, this implies a knowledge of ($i$) the contraction
time scale, which is still debatable (e.g. Palla \& Stahler 2002;
Hartmann 2003),  ($ii$) the probability for each element
to stick onto a dust grain (the so--called sticking coefficient, 
which is typically assumed of order unity, but see Jones \& Williams 
1985); (iii) the fraction of charged grains 
and their interaction with positive ions (i.e. all those elements with 
ionization potential less than that of H in diffuse clouds, such as
C, S, Si, Fe, Na, Mg; e.g. Ruffle et al. 1999; Weingartner \& Draine 2001).

Once the set of initial abundances has been assumed, one needs to 
solve the ``rate equations'', which govern the time variation 
of species both in the gas--phase and on the surface of dust grains (see e.g.
Hasegawa et al. 1992). This 
involves quite often the assumption of uncertain rate coefficients 
for gas--phase reactions and the poorly known processes happening 
on grain surfaces.  Nevertheless, chemical models are now reasonably 
able to reproduce observational data, thanks to the tight interaction 
between theorists, observers and experimentalists, fundamental for constraining 
models on the one side and guiding observations on the other. 
The next two subsections will treat gas--phase and surface 
chemistry separately, pointing out the basic concepts in both fields.

\begin{table}
\centering
\caption{Measured elemental abundances (with respect 
to total hydrogen) and those adopted in chemical models.}
\label{tab:element}       
%
%
\begin{tabular}{l|cc|cc}
\hline\noalign{\smallskip}
\multicolumn{1}{c|}{Element} & \multicolumn{2}{c|}{MEASURED} &
\multicolumn{2}{c}{ADOPTED} \\
& \multicolumn{1}{c}{~~Solar~~} & 
\multicolumn{1}{c|}{~~Cosmic~~} & \multicolumn{1}{c}{~High Metal} & 
\multicolumn{1}{c}{Low Metal}  \\
\noalign{\smallskip}\hline\noalign{\smallskip} 
C &  3.6(-4) & 2.1(-4) & 7.3(-5) & 7.3(-5) \\
N &  9.3(-5) & 6.6(-5) & 2.1(-5) & 2.1(-5) \\
O &  7.4(-4) & 4.6(-4) & 1.8(-4) & 1.8(-4) \\
S &  1.9(-5) & 1.2(-5) & 8.0(-6) & 8.0(-8) \\
Si & 3.6(-5) & 1.9(-5) & 8.0(-7) & 8.0(-9) \\
Fe & 3.2(-5) & 2.7(-5) & 3.0(-7) & 3.0(-9) \\
Na & 1.9(-6) & ...     & 2.0(-7) & 2.0(-9) \\
Mg & 3.8(-5) & 2.5(-5) & 7.0(-7) & 7.0(-9) \\
\noalign{\smallskip}\hline
\multicolumn{5}{l}{Note: a(b) $\equiv$ a$\times$10$^b$.} 
\end{tabular}
\end{table}

\subsection{Gas--phase chemistry}
\label{sec:gasphase}

The rate coefficient $k$(cm$^3$/s) of a generic reaction 
A+B$\rightarrow$C+D is given by:
\begin{eqnarray}
k & = & < \sigma \, v > 
\label{eq:k}
\end{eqnarray}
where $\sigma$ is the total cross section of the reactants, 
$v$ is the relative
velocity, and the averaging is performed over
the thermal distribution.  Most reactions have appreciable activation
energy ($E_{\rm a}$) 
even if exothermic (this is because chemical reaction typically 
involve the breaking of chemical bonds before the formation 
of new ones), and the rate coefficients $k$ can be expressed
by the simple Arrhenius formula (e.g. Herbst 1990): 
\begin{eqnarray}
k & = & A(T) \exp(-E_{\rm a} / k_{\rm B} T) 
\end{eqnarray}
where $k_{\rm B}$ is the Boltzmann constant, $T$ is the gas kinetic
temperature and $A(T)$ is the ``pre--exponential factor'',
a weak function of the temperature, which depends upon the actual shape of 
the reaction potential surface.
Activation energies of about 0.1--1 eV are normal, so that in 
dark clouds, where $k_{\rm B} T$ $\sim$ 0.01 eV, 
 even exothermic reactions may not proceed.  

Particularly important for the chemistry 
of dark clouds are ion--molecule reactions (Herbst \& Klemperer
1973), which do not possess activation energy 
because of the strong long--range attraction force, 
given by the potential:
\begin{eqnarray}
V(R) & = & - \frac{\alpha \, e^2}{2 \, R^4}  
\end{eqnarray}
where $\alpha$ is the polarizability of the neutral species (cm$^{3}$), 
$e$ is the electronic charge in esu, and $R$ is the distance between 
the two species.  For a sufficiently close encounter, so that 
orbiting or spiraling occurs ($R$ $\le$
($4 e^2 \alpha / \mu v^2)^{1/4}$, where $\mu$ is the reduced mass
in the collision), the rate coefficient becomes:                       
\begin{eqnarray}
k_{\rm L} & = &  2 \pi e \left( \frac{\alpha }{\mu} \right)^{1/2} \,
\simeq \, 10^{-9} \, {\rm cm^3} {\rm s^{-1}} , 
\end{eqnarray}
the so--called Langevin rate, 
independent of the temperature (Herbst 1990). 

The main processes happening in the gas phase are listed in 
Tab.~\ref{tab:gasphase} along with their typical rate coefficients
(see Tielens 2005 and Duley \& Williams 1984 for details). 
Here it is worthwhile mentioning some examples, 
important for dark clouds and star forming regions.

\begin{table}
\centering
\caption{Classes of gas--phase reactions}
\label{tab:gasphase}       
%
%
\begin{tabular}{lll}
\hline\noalign{\smallskip}
Type & Process & Rate  \\
        &       & Coefficient \\
\noalign{\smallskip}\hline\noalign{\smallskip} 
Ion--molecule & A$^+$ + B $\rightarrow$ C$^+$ + D & $\sim$10$^{-9}$ 
cm$^3$ s$^{-1}$ \\
Dissociative Recombination  & AB$^+$ + e $\rightarrow$ A + B & $\sim$10$^{-6}$
cm$^3$ s$^{-1}$ \\
Radiative Association & A + B $\rightarrow$ AB + {\it h $\nu$} & 
$\sim$10$^{-16}$--10$^{-9}$ cm$^3$ s$^{-1}$ \\
Neutral--neutral & A + B $\rightarrow$ C + D & $\sim$10$^{-12}$--10$^{-10}$
 cm$^3$ s$^{-1}$\\
Photodissociation & AB + $h \nu$ $\rightarrow$ A + B & $\sim$10$^{-9}$ 
s$^{-1}$ \\
Charge--transfer & A$^+$ + B $\rightarrow$ A + B$^+$ & $\sim$10$^{-9}$ 
cm$^3$ s$^{-1}$ \\
\noalign{\smallskip}\hline
\multicolumn{3}{l}{NOTE: Photodissociation rate listed for no extinction only.}
\end{tabular}
\end{table}

\noindent
{\it Ion--molecule reactions and dissociative recombination} \\
The production of ions in dark molecular clouds is initiated by cosmic rays
($c.r.$),
which ionize H$_2$, the most abundant molecular species.  The rate 
of this process, called $\zeta$, is typically of order 
10$^{-17}$ s$^{-1}$ (e.g. van der Tak \& van Dishoeck 2000), 
and about 97\% of the ionizations produce H$_2^+$:
\begin{eqnarray}
{\rm H_2} + c.r. & \rightarrow & {\rm H_2^+} + e + c.r. 
\end{eqnarray}
Once H$_2^+$ is formed, it reacts very quickly with H$_2$ to produce 
H$_3^+$, one of the most important molecular ions in astrochemistry:
\begin{eqnarray}
{\rm H_2^+} + {\rm H_2} & \rightarrow & {\rm H_3^+} + {\rm H}
\end{eqnarray}
In general, if an ion can react with H$_2$, no other reaction need
to be considered because H$_2$ is at least three orders of magnitude 
more abundant than the other species and the reaction rate is proportional
to the abundance of the reactants.  H$_3^+$ does not react with H$_2$ but it can 
easily transfer a proton, thus increasing molecular complexity. 
For example, if H$_3^+$ encounters an atomic oxygen, OH$^+$ is produced
(${\rm H_3^+} + {\rm O} \rightarrow {\rm OH}^+ + {\rm H_2}$).
Then, successive ``H atom'' transfer (ion--H$_2$) reactions 
proceed until a saturated species (which cannot further react 
with H$_2$) is formed: 
OH$^+$ $\stackrel{\rm H_2}{\longrightarrow}$ H$_2$O$^+$ 
$\stackrel{\rm H_2}{\longrightarrow}$ H$_3$O$^+$. Upon {\it dissociative 
recombination} of the oxonium ion H$_3$O$^+$,  water and OH form:
\begin{eqnarray}
{\rm H_3O^+} + e & \rightarrow & {\rm H_2O} + {\rm H} \\
		 & \rightarrow & {\rm OH} + 2 {\rm H} \,\, {\rm or} \,\, 
		  {\rm OH} + {\rm H_2} 
\end{eqnarray}
The fraction of H$_3$O$^+$ ions which will produce H$_2$O and OH upon 
dissociative recombination is called the {\it branching ratio} and 
it is measured in the laboratory (e.g. Geppert et al. 2004).  
Ion--molecule chemistry of larger species leads to 
mainly H--poor (unsaturated) species, unlike dust chemistry (see 
Sec.~\ref{sec:surface}).

\noindent
{\it Neutral--neutral reactions} \\
Although they typically possess activation energy (and thus
become important in high temperature regions; e.g. Sect.~\ref{sec:outflows}), 
some neutral--neutral reactions are especially crucial in dark clouds
for nitrogen and sulphur chemistry,  which starts with: 
\begin{eqnarray}
{\rm CH} + {\rm N} & \rightarrow & {\rm CN} + {\rm H} \\
{\rm S} + {\rm OH} & \rightarrow & {\rm SO} + {\rm H} 
\end{eqnarray}
The reaction of CN with N will produce N$_2$ (also formed by NO + N), 
which can react with H$_3^+$ to form the easily observable N$_2$H$^+$.  N$_2$
initiates the chemical path toward the production of another
well known interstellar molecule: NH$_3$ (N$_2$ first reacts with
He$^+$ to produce N$^+$, which is then successively hydrogenated 
by H$_2$ until the saturated NH$_4^+$ ion is formed; NH$_3$
is finally produced via NH$_4^+$ dissociative
recombination).  The common chemical origin between 
N$_2$H$^+$ and NH$_3$ explains the good correlation between the 
two species found in molecular cloud cores (Benson et al. 1998).

\subsection{Surface chemistry}
\label{sec:surface}

The fundamental role played by dust grains in synthesizing 
molecular species is well known for about fourty years, 
when they were invoked to explain the production rate of 
H$_2$ molecules in our Galaxy.  Pioneer work in this field
has been done by Gould \& Salpeter (1963) and Hollenbach \& Salpeter (1971), 
who first derived the formation rate of H$_2$ on grain surfaces.
From observations carried out by the Copernicus satellite,
the derived rate of formation for H$_2$ in diffuse clouds
is $R$ $\simeq$ 3$\times$10$^{-17}$ cm$^{3}$s$^{-1}$ (Jura 1975).    
Lots of theoretical and laboratory work has been done since then 
on this fundamental process (e.g. Pirronello et al. 1999;  
 Williams et al. 2000; Cazaux \& Tielens 2004). 
Observations of the so--called HINSA (HI Narrow Self--Absorption) in 
cloud cores are also giving clues on the  
H$_2$ formation efficiency in regions where dust grains are covered
by icy mantles (e.g. Li \& Goldsmith 2005). 

Surface chemistry is also invoked to explain ($i$) the observed large 
abundances of complex species near star forming 
regions (Sect.~\ref{sec:hotcore});  ($ii$) the orders--of--magnitude 
abundance enhancements of selected molecular species  
observed along molecular outflows 
(Sect.~\ref{sec:outflows}); ($iii$) dust grain mantles composed 
of complex icy mixtures, as deduced from ground--based and space
-- in particular, ISO, the Infrared Space Observatory -- observations 
of background field stars or embedded protostars
(e.g. Sellgren et al. 1994; 
Gibb et al. 2004; van Dishoeck 2004); ($iv$) the large amount 
of deuterium fractionation  observed in low mass 
star forming regions (Sect.~\ref{sec:prestellar} and 
~\ref{sec:hotcore}).  

\begin{figure}
\centering
\includegraphics[height=10cm,angle=270]{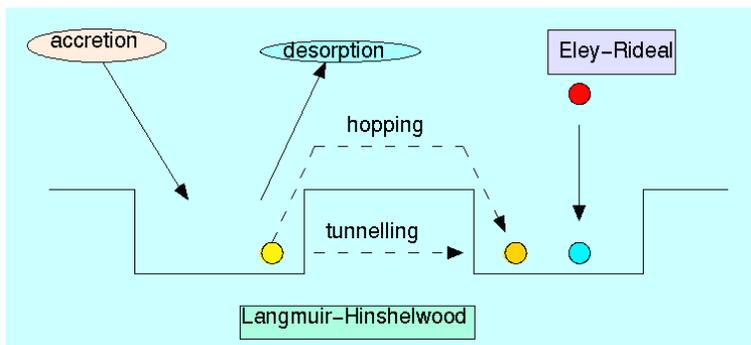}
%
%
\caption{Assorted processes on dust grain surfaces (from Herbst 2000).}
\label{fig:surface}       
\end{figure}

The main surface processes are illustrated in Fig.~\ref{fig:surface}
(from Herbst 2000), which shows a portion of an idealized grain 
surface, with  adjacent sites. Once a 
light atom (such as H) hits the dust surface, it is first 
 {\it adsorbed} in a binding site.  The species is 
{\it physisorbed} if it is linked to the surface by 
van der Waals forces,  and  binding energies are in the range 0.01--0.2 
eV (100--2000 K).  In this case, light species can quickly ($<<$ 1 s) 
move across the surface, overcoming the energetic barriers via 
thermal hopping or quantum tunneling, 
depending upon the dust temperature (e.g. Tielens \& Allamandola 1987). 
Indeed, surface chemistry is not 
limited by reaction rates but by the {\it accretion rates} of reactive species 
on grain surfaces (e.g. in a typical dark cloud, one H atom hits one 
dust grain approximately once every 10 days). 

Surface chemistry 
based on diffusion is known as a {\it Langmuir--Hinshelwood} (LH) chemistry
(see also Tielens \& Hagen 1982).  Here, {\it hydrogenation} is one of the 
main surface processes, so that O is transformed to H$_2$O (the main 
ice component), C to CH$_4$, 
N to NH$_3$, S to H$_2$S,  and CO to H$_2$CO and 
CH$_3$OH.  Hydrogenation of CO implies the overcoming of an activation
energy barrier (Woon 2002),
which slows down the process;  in fact, CO is one of the major 
compounds of grain mantles observed in molecular clouds, second only to water  
(e.g. Chiar et al. 1995; Teixeira et al. 1998).  There are still some 
doubts about the existence of 
solid NH$_3$ (e.g. Schutte et al. 2003) and H$_2$S (see discussion in Wakelam et al. 2004).  
But solid OCS has been detected (Palumbo 
et al. 1997) and CO$_2$ absorption features are ubiquitous in the direction of 
background and embedded objects
(e.g. Nummelin et al. 2001).  Therefore, oxygenation is 
likely to be another important surface process, which 
competes with hydrogenation depending on the gaseous O/H ratio and the 
dust temperature (H being more volatile than O).

%

If the accreted species are strongly bound, or {\it chemisorbed}, with 
typical strengths of $\sim$ 1 eV (10000 K), the {\it Eley--Rideal}
(ER) mechanism dominates (see Fig.~\ref{fig:surface}):  
a surface reaction only occurs when a gas--phase species 
lands atop or nearby the chemisorbed species. 
This type of chemistry, efficient only for high surface
coverage, preferentially occurs at dust temperatures
$T_{\rm d}$ below about 300 K (for a barrier between chemisorption
sites of $\sim$1~eV, S. Cazaux, priv. comm.), 
when the LH mechanism cannot proceed 
for chemisorbed species (see discussion in Habart et al. 2004 on the
importance of the various mechanisms for H$_2$ formation in 
photodissociation regions).  

The return of surface species back in the gas phase can be driven 
by {\it thermal desorption}, with a time scale exponentially dependent
on the dust temperature. At $T_{\rm d}$ = 10 K,  only H and D can 
thermally evaporate with significantly large rates, whereas 
$T_{\rm d}$ $>$ 90 K is needed to start H$_2$O evaporation.  
Thus, large abundances of water 
and other saturated species are expected
in the gas phase in star forming regions where the dust has been 
heated above, say, 100 K by the protostellar radiation field 
(see Sect.~\ref{sec:hotcore}).  Among the main {\it nonthermal desorption} 
mechanisms, important in cold clouds, are: ($i$) cosmic--ray bombardment 
(especially by the heavier nuclei; L\'eger et al. 1985; 
Hasegawa \& Herbst 1993), and ($ii$) the energy
generated by chemical surface reactions, still debatable
 (Willacy \& Millar 1998; Takahashi \& Williams 2000).   
UV photons play a role in photodesorption
of solid species in PDRs and diffuse clouds (although the yield 
of photodesorption is still not well constrained; e.g. Ruffle \& Herbst 2001). 

\section{Astrochemistry in star forming regions}
\label{sec:star}

In this section, the main chemical processes in some of the 
regions of interest for star formation will be reviewed.  For 
a more detailed description of the physical properties of 
molecular cloud cores the reader is encouraged to read 
Walmsley (this book), Myers (this book) and Tafalla (this book).

\subsection{Pre--stellar cores}
\label{sec:prestellar}

Stars form within the densest portions of molecular clouds, the 
``dense cores'' (e.g. Myers 1999), upon gravitational collapse.  Before the 
formation of a protostellar object, (starless) dense cores 
are characterized by low temperatures ($\sim$ 10 K) and 
relatively high central densities ($\simeq$ 10$^5$ cm$^{-3}$).  
These conditions favour the freeze--out of gas phase species onto 
dust grains.  This can be easily understood, comparing the freeze 
out and the dynamical time scales. The 
time for a gaseous species X to accrete onto a dust grain is given by:
\begin{eqnarray}
t_{\rm fo} & = & \frac{1}{S n_{\rm d} \pi a_{\rm d}^2 v_{\rm t}} \,\,  \\
           & \simeq & 10^9 \sqrt{m_{\rm X}/T} (n_{\rm H} S)^{-1} \,\, {\rm yr} ,
\end{eqnarray}
where S is the sticking coefficient or the probability of a certain 
species to stick on the grain upon collision 
(about unity), $n_{\rm d}$ is the number
density of dust grains, $a_{\rm d}$ is the grain radius, $v_{\rm t}$
is the thermal velocity of the gaseous species, $n_{\rm H}$ is 
the number density of hydrogen nuclei (it has been assumed that 
the gas--to--dust density ratio (by mass) is 100 and $a_{\rm d}$ = 0.1 
$\mu$m), and $m_{\rm X}$ is the mass of species X in amu.  
Therefore, in a cloud with number density $n_{\rm H}$ = 
10$^5$ cm$^{-3}$, $t_{\rm fo}$ $\sim$ 10$^4$ yr. 
The free--fall time scale, the shortest dynamical time
scale, for a cloud with the same $n_{\rm H}$ value is (e.g. Spitzer 1978):
\begin{eqnarray}
t_{\rm ff} & = & 4\times 10^{7} (n_{\rm H})^{-1/2} \,\, (\rm yr),
\end{eqnarray} 
or 10$^5$ yr.  Thus, freeze out is expected to be 
an important process in the evolution of cloud cores toward the 
formation of a star.  

\begin{figure}
\centering
\includegraphics[height=10cm,angle=270]{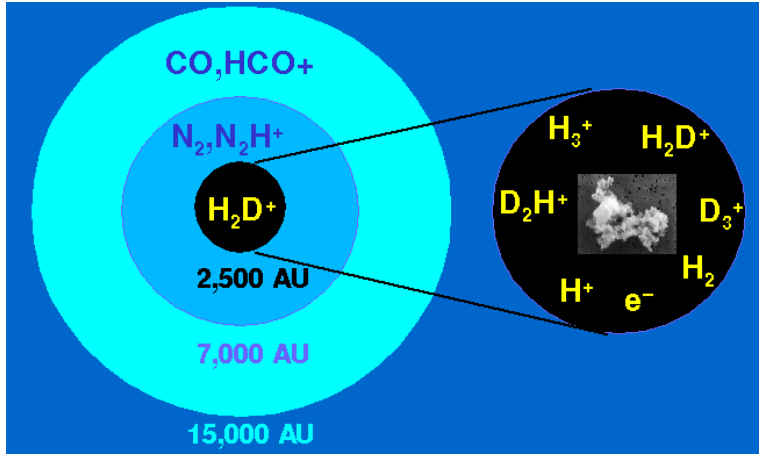}
%
%
\caption{Schematic picture of the chemical structure of a typical pre--stellar 
core.  The central density is around 10$^6$ cm$^{-3}$ and it
drops to $\sim$2$\times$10$^4$ cm$^{-3}$ at the edge.
In the outer shell, CO is mostly in the gas phase and HCO$^+$ is the main ion.
Within about 7,000 AU, CO is mainly frozen onto 
dust grains whereas N$_2$ remains in the gas phase, so that
a good probe of these zones is N$_2$H$^+$. In the central 
$\sim$ 2,500 AU (the ``molecular hole''), only light species survive 
in the gas phase.  Here, the thick icy
mantle may boost the coagulation of dust grains.}
\label{fig:prestellar}       
\end{figure}

In fact, strong depletion of CO from the gas phase has been observed 
in a variety of molecular cloud cores (e.g. Caselli et
al. 1999; Bergin et al. 2001; Tafalla et al. 2004).
Tafalla (this volume) presents a detailed review of the observed chemical properties
of starless cores.  Here, I summarize the main observed properties of 
{\it pre--stellar cores}, 
which differ from starless cores in being more centrally concentrated and
showing kinematic evidences of central infall (Ward--Thompson et al. 
1999; Crapsi et al. 2005). Pre--stellar 
cores are thought to be unstable objects, on the verge of star formation, 
thus their study is crucial to unveil the initial conditions
of the star formation process. Fig.~\ref{fig:prestellar} shows a schematic picture 
of the chemical structure of a ``typical'' pre--stellar core: 
($i$) the core envelope ($\sim$ 7,000--15,000 AU):  CO is mainly in the gas 
phase, and the main molecular 
ion is HCO$^+$; ($ii$) the core nucleus ($\sim$ 2,500--7,000 AU):  CO 
and other C--bearing species are highly depleted from the gas phase, unlike
N--bearing species (e.g. Bergin \& Langer 1987); thus,
N$_2$H$^+$ is a good tracer of these zones and  
the deuterium fractionation is very large (see below); 
($iii$) the ``molecular hole'' (within $\sim$ 2,500 AU): all 
species heavier than helium are likely to be condensed out onto dust grains
(the best tracer of this zone may be H$_2$D$^+$). 

%

\noindent
{\it Freeze--out and deuterium fractionation} \\
An important consequence of the depletion of gas phase species is the 
large enhancement in the deuterium fractionation of molecular species, which 
has been extensively observed in the past years (e.g. Caselli et al. 2002;
van der Tak et al. 2002; Bacmann et al. 2003).  This process involves the
deuteron--proton exchange reaction (Watson 1974):
\begin{eqnarray}
{\rm H_3^+} + {\rm HD} \rightarrow {\rm H_2D^+} + {\rm H_2} + \Delta {\rm E} ,
\label{eq:h2dp}
\end{eqnarray} 
with $\Delta$E $\simeq$ 230 K, which can only proceed 
from left to right in cold dark clouds.  Here, in fact, the 
${\rm H_2D^+}/{\rm H_3^+}$ abundance ratio increases well above the 
cosmic D/H ratio ($\sim$ 1.5$\times$10$^{-5}$).
 ${\rm H_2D^+}$ can react with CO, N$_2$ and other neutrals, 
transferring the deuteron in at least one third of its encounters,
with the consequence of enhancing the deuterium fractionation in, e.g., 
HCO$^+$ and ${\rm N_2H^+}$, as observed
(e.g. Williams et al. 1998).  In the case of 
CO (and other ${\rm H_3^+}$ and ${\rm H_2D^+}$ destruction partners) depletion, 
the ${\rm H_3^+}$ abundance will increase and, consequently,
the ${\rm H_2D^+}$ production rate will raise (via reaction ~\ref{eq:h2dp}),
further increasing the ${\rm H_2D^+}/{\rm H_3^+}$ abundance ratio 
and the deuterium fractionation.  In pre--stellar cores, 
${\rm N_2D^+}/{\rm N_2H^+}$ $\simeq$ 0.2 (Caselli et al. 2002; Crapsi et al.
2005) and ${\rm D_2CO}/{\rm H_2CO}$ $\simeq$ 0.05 (Bacmann et al. 2003), orders 
of magnitude larger that the cosmic D/H ratio.  The observed 
D--fractionation also correlates with the measured amount of CO 
depletion (Bacmann et al. 2003; Crapsi et al. 2005), as expected
from theory (e.g. Dalgarno \& Lepp 1984; 
Aikawa et al. 2005).

In {\it molecular holes}, volume densities are large enough
to allow an ``extreme'' depletion of gaseous species heavier than helium.  
This seems to be supported by the detection of a strong ${\rm H_2D^+}$ line 
in the direction of L1544 (Caselli et al. 2003). The estimated ${\rm H_2D^+}$ 
abundance cannot be reproduced by standard gas--grain chemistry unless 
CO is mainly in solid form (e.g. Roberts et al. 2003; Walmsley et al. 
2004). If this is the case, the only tracer of the chemical composition 
of this region may be ${\rm H_2D^+}$. Also ${\rm D_2H^+}$ has been 
observed in another pre--stellar core (Vastel et al. 2004), with 
abundances similar to ${\rm H_2D^+}$, which further pointed out 
the importance of including  multiply deuterated species in chemical 
models of pre--stellar cores. 

\subsection{Outflows}
\label{sec:outflows}

 In their youngest phase, protostars
are called Class 0 sources (Andr\'e et al. 1993) and one of their main
characteristics is the presence of powerful and collimated outflows 
(e.g. Bachiller 1996; see also Beuther \& Shepherd, 
this book and Bacciotti, this 
book). Outflows shock the material along their path, raising 
the gas temperature and partially destroying dust grains.  In the 
presence of magnetic fields, if the electron fraction is sufficiently 
low (as in the dense material surrounding Class 0 sources), 
and if the shock velocity is below about 50 km s$^{-1}$ 
(Draine et al. 1983), a discontinuity in the 
hydrodynamical variables (e.g. density and temperature) is not 
present and the shock is named ``C--type'' (``C'' stays for ``continuous''). 
  In all the other cases, a ``jump front'' is formed, and
the shock is called ``J--type'' (see Draine 1980).

In ``C--type'' shocks, molecular species (in particular H$_2$) do not dissociate. 
In these conditions, and if the gas temperature raises above $\sim$200 K, 
some important endothermic reactions become fast and quickly convert most of 
the free oxygen into water (${\rm O} \stackrel{\rm H_2}{\longrightarrow} 
{\rm OH} 
\stackrel{\rm H_2}{\longrightarrow} {\rm H_2O}$),
and sulphur mainly into SO and SO$_2$ (e.g. Hartquist et al. 1980; Pineau 
des For\^ets et al. 1993).
  In dissociative shocks, as soon as 
the post--shocked gas temperature has fallen to a few hundred degrees, 
H$_2$ molecules are reformed and shock chemistry proceeds 
as in C--type shocks (e.g. Neufeld \& Dalgarno 1989). 

The presence of magnetic fields in C--type shocks implies 
different velocities for neutrals and charged particles, including
dust grains which are negatively charged and linked to the magnetic 
field lines. In these conditions, the collisions between 
(especially) H$_2$ molecules and dust grains may cause the return of 
mantle and refractory species in the gas phase.  This process is called 
grain {\it sputtering} (e.g. Flower \& Pineau des For\^ets 1995; Schilke 
et al. 1997) and it is thought to be responsible for the observed 
enhancement of SiO and saturated 
molecular species such as H$_2$O, NH$_3$, and CH$_3$OH along 
protostellar outflows (e.g. 
Bachiller \& Perez Gutierrez 1997; J\o rgensen et al. 2004).  
Dust grains with different sizes will
also have relative velocities because of the different mass and charge
(entering the gyrofrequency expression),
so that grain--grain collisions are also quite 
efficient in vaporising mantle and refractory material (Caselli et al. 1997).  

Once icy mantles
and part of the refractory material are released in the gas phase, shock
chemistry proceeds until the gas temperature drops to 
the pre--shock values. This happens in a short time scale (a few hundred 
years), because of the efficient gas cooling mainly due to H$_2$, water 
and CO  (e.g. Kaufman \& Neufeld 1996).  Then, the post--shocked
gas will experience again the ``cold chemistry'', including
freeze--out (e.g. Bergin et al. 1998). This mechanism 
(shock+freeze--out) has been suggested by Bergin et al. (1999) as 
a way to explain the large abundance ($\sim$10$^{-4}$ w.r.t. H$_2$) 
of water in solid form in molecular clouds.  

\subsection{Hot Cores}
\label{sec:hotcore}

The newborn protostar eventually heats up 
the surrounding material with its radiation, raising the dust temperature
above the sublimation value for grain mantles, and forming the so--called
``Hot Cores'' (HCs). HCs are typically associated with high mass star forming
regions and represent a stage earlier than ultracompact HII regions. They 
have diameters of $<$0.1 pc, H$_2$ number densities
of $\ge$ 10$^7$ cm$^{-3}$ and gas temperatures $\ge$100 K (e.g. 
Kurtz et al. 2000).  Their chemical 
signatures are:  ($i$) large abundances of saturated species, such as 
H$_2$O, NH$_3$, H$_2$S, CH$_3$OH (e.g. Pauls et al. 1983; 
Menten et al. 1986), suggestive of an active surface chemistry 
(Sect.~\ref{sec:surface}); 
($ii$) large abundances of complex organic H--rich molecules, in particular 
HCOOCH$_3$ (methyl formate), CH$_3$COOH (acetic acid), 
CH$_3$CH$_2$CN (ethyl cyanide), CH$_3$CH$_2$OH
(ethanol), (CH$_3$)$_2$CO (acetone) (e.g. Millar et al. 
1995; Mehringer \& Snyder 1996; Remijan et al. 2004), 
hard to form in the gas phase; 
($iii$) relatively large deuterium fractionations (e.g. 
Oloffson 1984; Turner 1990), a record of an early cold phase
(Sect.~\ref{sec:prestellar}).
Interestingly, 
these regions show a strong chemical differentiation, especially 
evident in the nearby Orion KL (e.g. Blake et al. 1987;
Wright et al. 1996) and also in the W3(OH) region 
(Wyrowski et al. 1997), when millimeter interferometry is used.

To understand the gas phase composition in HCs, 
surface chemistry is needed, because gas phase routes appear to be too slow.
In particular,  saturated species can be efficiently formed on grain surfaces
before or during the earliest stages of protostar evolution, when the
dust temperature is low enough ($\le$ 20K) to allow hydrogenation  
and produce solid H$_2$O, NH$_3$, CH$_3$OH and H$_2$S 
(see Sect.~\ref{sec:surface}).  Once the dust is heated to 
temperatures larger than the sublimation values for icy mantles (about 
100 K), these saturated species will enrich the gas phase and start 
a hot gas phase chemistry (e.g. Brown et al. 1988; 
Charnley et al. 1992).  A way to reproduce the observed chemical differentiation
in the Orion KL region, where N--bearing and O--bearing species
are spatially segregated, is to take into account the 
temperature and density gradients of the circumstellar cloud 
(Caselli et al. 1993).  

In general, different routes are invoked
for complex O--bearing and N--bearing species: the former are produced
after the evaporation of solid methanol (CH$_3$OH) and formaldehyde 
(H$_2$CO) (the complexity is increasing in the gas phase); the latter 
are formed on grain surfaces (e.g. ethyl cyanide, CH$_3$CH$_2$CN)
and, once evaporated, they form simpler daughter species (e.g. 
vinyl cyanide, CH$_2$CHCN).
However, as recently found in the laboratory (Horn et al. 2005), 
some key reactions in the chain to form the HC--ubiquitous 
methyl formate from evaporated 
methanol do not proceed.  Thus, surface 
chemistry is also  needed to produce complex O--bearing species.   
{\it This is all to be explored!}
In fact, current chemical models neglect surface chemistry, although they
are very sophisticated in treating the physical structure of
the molecular clouds (e.g. Doty et al. 2002; Rodgers \& Charnley 2003; Nomura \& Millar 2004;
Viti et al. 2004).

In the past few years it has been realized that HCs are not only a 
signature of high mass star formation, but they are also found in  low mass 
star forming regions, during the earliest (Class 0)
protostellar phase (Cazaux et al. 2003; Bottinelli et al. 2004a,b; Kuan
et al. 2004), and in one intermediate mass Class 0 source (Fuente
et al. 2005).  Low--mass HCs, called ``hot corinos'', are a ``scaled'' 
version ($n({\rm H_2})$ $\sim$ 
10$^7$ cm$^{-3}$, $T$ $\sim$ 50--100 K, size $\sim$ 150 AU) of the 
massive HCs.  Nevertheless, they show
rich spectra at millimeter wavelengths, with complex organic molecules,
which is challenging current theory (see e.g. Rodgers \& Charnely 2001). 
Outside these ``hot corinos'', the gas and the dust
still maintain the pre--stellar--core characteristics, including 
freeze--out (e.g. Belloche \& Andr\'e 2004).  In the well--studied case
of IRAS~16293-2422, the deuterium fractionation is very large (especially
for H$_2$CO and CH$_3$OH; e.g. Parise et al. 2004) compared to high 
mass HCs, probably suggesting a different chemical evolution, such as
lower 
dust temperatures during the pre--stellar phase, which favour surface
deuteration, or different chemical time scales. 

\subsection{Protoplanetary Disks}
\label{sec:ppdisks}

Protoplanetary disks around T Tauri and Herbig Ae/Be stars, which are
contracting toward the main sequence and have ages of a few hundred years, 
are an important link between molecular clouds 
and protoplanetary systems (see Markwick \& Charnley 2004 for a recent
review).  Understanding their chemical composition
and the chemical processes is in fact one of the major challenge 
for astrochemistry, although the next generation of telescopes (in 
particular ALMA, the Atacama Large Millimeter Array) is needed to 
put stringent constraints on the models. So far, observations 
of species such as CO, CS, CN, HCN, HCO$^+$, N$_2$H$^+$, and H$_2$CO 
have shown that molecular abundances are typically lower than 
those in dense interstellar clouds, suggesting that molecular freeze--out
is at work (e.g. Dutrey et al. 1997; Goldsmith et al. 1999; 
Qi et al. 2003; Thi et al. 2004).  
Moreover, DCO$^+$ has also 
been detected and the derived deuterium fractionation is comparable 
to that observed in dense cloud cores (DCO$^+$/HCO$^+$ $\sim$ 0.04; 
van Dishoeck et al. 2003). Ceccarelli et al. (2004) have detected 
H$_2$D$^+$, known to be abundant in highly depleted regions 
(see Sect.~\ref{sec:prestellar}). There is also observational 
evidence that photon--dominated chemistry is significant (because 
of the relatively large abundance of CN and C$_2$H, e.g. Dutrey 
et al.  1997).  Theoretical work has pointed out the importance
of X--rays in the ionization (e.g.  Glassgold et al. 1997) and 
thermal (e.g. Glassgold et al. 2004) balance.   Thus, many ingredients
participate in the chemistry of protoplanetary disks.

\begin{figure}
\centering
\includegraphics[height=10cm,angle=270]{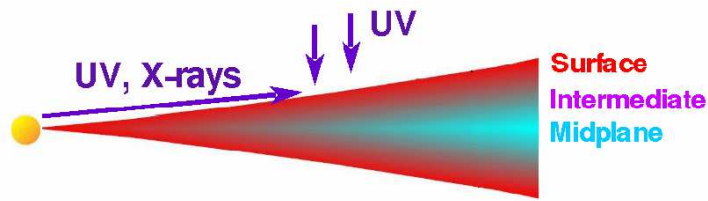}
%
%
\caption{Schematic picture of a protoplanetary disk, where three chemically
distinct zones can be recognized: the {\it midplane}, where freeze--out is 
dominant; the {\it intermediate zone}, where dense cloud chemistry 
is appropriate; the {\it surface}, where photochemistry is important (courtesy
of Y. Aikawa).}
\label{fig:ppd}       
\end{figure}

In current chemical models, protoplanetary disks can be divided in 
three layers (see Fig.~\ref{fig:ppd}):
the {\it surface layer}, where photochemical processes are important
because of the exposure to the UV flux from the star and the 
interstellar radiation field; the {\it intermediate layer}, where 
the extinction is large enough that the chemistry resembles that of 
dense clouds; the {\it midplane}, where the density is so large 
that the majority of gas phase species are frozen onto dust grains
(see, e.g., Aikawa et al. 2001; Willacy \& Langer 2000; 
Markwick et al. 2002; van Zadelhoff et al. 2003; Semenov et al. 2004). 

\section{Conclusions}
\label{sec:conclusion}

Tab.~\ref{tab:summary} summarizes the main processes for the 
chemistry of star forming regions, in the three phases (pre--stellar
cores, Class 0 protostars 
and protoplanetary disks) briefly illustrated in this review.
Question marks in the table follow not--yet--proved statements, 
mostly based on model predictions.     

\begin{table}
\centering
\caption{Main processes and molecular gas tracers in star forming
regions}
\label{tab:summary}       
%
%
\begin{tabular}{lll}
\hline\noalign{\smallskip}
Region & Gas Tracer & Main Processes \\
\noalign{\smallskip}\hline\noalign{\smallskip} 
{\it Pre--stellar cores}: &  \\
envelope & CO, HCO$^+$, CS & ion--molecule \\
nucleus & ${\rm N_2H^+}$, ${\rm N_2D^+}$, NH$_3$ & freeze--out, 
D--fractionation \\
molecular hole & ${\rm H_2D^+}$, ${\rm D_2H^+}$, D$_3^+$? & freeze--out,
grain coagulation? \\
\noalign{\smallskip}\hline
{\it Class 0}: & \\
envelope & ${\rm N_2H^+}$, NH$_3$ & ion--molecule, freeze--out \\
outflow & H$_2$O, SiO, CH$_3$OH, SO$_2$ & grain sputtering, 
neutral--neutral \\
hot cores & H$_2$O, complex organics & mantle evaporation,
neutral--neutral \\
\noalign{\smallskip}\hline
{\it Protoplanetary disks}: & \\
surface layer & CO, CN, C$_2$H  & photochemistry, ion--molecule \\
intermediate layer & HCN, DCO$^+$, N$_2$H$^+$ & freeze--out, D--fractionation \\
midplane & H$_2$D$^+$, H$^+$?, D$_2$H$^+$?, D$_3^+$? & freeze--out, 
grain coagulation \\ 
\noalign{\smallskip}\hline
\end{tabular}
\end{table}

Here, I would like to point out some of the main problems 
still open in the field of astrochemistry of star forming regions. 
Many uncertainties are present especially 
regarding surface chemistry (e.g. Stantcheva \& Herbst 2004; Chang et al.
2004) and gas phase 
reactions at high temperature (e.g. Wakelam et al. 2004), which in particular
affect our interpretation of the observations toward molecular 
outflows and hot cores. Some important problems to solve are:
($i$) after recent laboratory work (Horn et al. 2004; Luca et al. 2004), 
we are left with the non--trivial problem of producing 
HCOOCH$_3$ and CH$_3$OH in the gas phase.  It seems that the formation
on grain surfaces 
(not yet explored for methyl formate) followed by desorption is the only 
possible route.  ($ii$) Grain mantle composition
is expected to change during cloud evolution and contraction; thus, binding 
energies (and consequently freeze--out rates) are a function of time.  
Nevertheless, current models simply neglect this.  ($iii$) There are still 
quite large uncertainties regarding the chemistry of Oxygen (e.g. how much 
atomic O is left in the gas phase in pre--stellar core?), and Sulphur 
(where does S go in dense clouds and what is its form in icy grain mantles?). 
($iv$) How good is the approximation of assuming a constant cosmic ray ionization 
rate ($\zeta$) throughout the core? A large value of $\zeta$ ($\sim$ 10$^{-15}$
s$^{-1}$) has been deduced for diffuse clouds (McCall et al. 2003). But 
this value drops to 1--3$\times$10$^{-17}$ s$^{-1}$ 
in dense clouds; does it continue to drop {\it within} pre--stellar cores, 
with a consequent variation of the fractional ionization and the chemistry 
in cloud core nuclei? A synergy of laboratory, theoretical, and 
observational work, as well as the new generation of telescopes and 
interferometers (ALMA, SMA, CARMA, Herschel, SOFIA, APEX), are sorely needed 
to advance our understanding of astrochemistry. 

\vspace{0.5cm}

\noindent
{\bf Acknowledgements} 

I thank all my collaborators, who keep feeding my enthusiasm about 
astrochemistry.  
A special thank to Eric Herbst and Malcolm Walmsley to have had the 
patience of going through 
this review and giving me important input.  Finally, I thank all the 
organizers of this meeting and, in particular, Nanda Kumar for all his 
hard work, needed to make the {\it C2C} a very successful (and enjoyable) 
meeting.


%
%
%
%
%
%
%
%
%
%
%
%
%
%
\input{referenc}



\printindex
\end{document}

%% file: referenc.tex
%
%

%
%

%% file: pcaselli.bbl
\begin{thebibliography}{99.}
%
%
%





\bibitem{} Aikawa, Y., Herbst, E.  (2001). A\&A, 371, 1107

\bibitem{} Aikawa, Y., Herbst, E., Roberts, H., Caselli, P.
(2005). ApJ, 620, 330

\bibitem{} Anders, E., Grevesse, N. (1989).
Geochimica et Cosmochimica Acta, 53, 197

\bibitem{} Andr\'e, P., Ward-Thompson, D. Barsony, M. (1993).
ApJ, 406, 122

\bibitem{} Asplund, M., Grevesse, N., Sauval, J. (2005).
In: \textit{Cosmic abundances as records of stellar evolution and 
nucleosynthesis} ed by F.N. Bash \& T.G Barnes (ASP conf. series), 
in press (astro-ph/0410214)

\bibitem{} Bachiller, R.  (1996). ARA\&A, 34, 111

\bibitem{} Bachiller, R., Perez Gutierrez, M. (1997). ApJ, 487, L93

\bibitem{} Bacmann, A., Andr\`e, P., Puget, J.-L., Abergel, A., 
Bontemps, S., Ward-Thompson, D.  (2000). A\&A, 361, 555 

\bibitem{} Bacmann, A., Lefloch, B., Ceccarelli, C.,
    Steinacker, J., Castets, A., Loinard, L. (2003). ApJ, 585, L55

\bibitem{} Bell, M. B., Matthews, H. E. (1985). ApJ, 291, L63

\bibitem{} Belloche, A., Andr\'e, P.  (2004). A\&A, 419, L35

\bibitem{} Benson, P. J., Caselli, P., Myers, P. C. (1998). ApJ, 506,
743

\bibitem{} Bergin, E. A., Ciardi, D. R., Lada, C. J., Alves, J.,
Lada, E. A. (2001). ApJ, 557, 209

\bibitem{} Bergin, E. A., Langer, W. D. (1997). ApJ, 486, 316

\bibitem{} Bergin, E. A., Neufeld, D. A., Melnick, G. J. 
(1998). ApJ, 499, 777

\bibitem{} Bergin, E. A., Neufeld, D. A., Melnick, G. J. (1999).
ApJ, 510, L145

\bibitem{} Bianchi, S., Gon\c alves, J., Albrecht, M., Caselli, P., 
Chini, R., Galli, D., Walmsley, C. M. (2003). A\&A, 399, L43

\bibitem{} Blake, G.A., Sutton, E.C., Masson, C.R., Phillips, T.G.
 (1987). ApJ, 315, 621

\bibitem{} Bottinelli, S., Ceccarelli, C., Lefloch, B.,  et al.  
(2004). ApJ, 615, 354

\bibitem{} Bottinelli, S., Ceccarelli, C., Neri, R., et al.  (2004). 
ApJ, 617, L69

\bibitem{} Brown, P. D., Charnley, S. B., Millar, T. J.
(1988). MNRAS, 231, 409

\bibitem{} Caselli, P., Hartquist, T. W., Havnes, O.  (1997). A\&A, 
322, 296

\bibitem{} Caselli, P., Hasegawa, T. I., Herbst, Eric (1993). ApJ, 408, 548

\bibitem{} Caselli, P., van der Tak, F. F. S.,
      Ceccarelli, C., Bacmann, A. (2003). A\&A, 403, L37

\bibitem{} Caselli, P., Walmsley, C. M., Tafalla, M.,
    Dore, L., Myers, P. C. (1999). ApJ, 523, L165

\bibitem{} Caselli, P., Walmsley, C. M., Zucconi, A., Tafalla, M., Dore, L., 
Myers, P. C. (2002). ApJ, 565, 344

\bibitem{} Ceccarelli,C., Dominik,C., Lefloch,B., Caselli,P., Caux, E.
(2004). ApJ,607,L51

\bibitem{} Cazaux, S., Tielens, A. G. G. M. (2004). ApJ, 604, 222

\bibitem{} Cazaux, S., Tielens, A. G. G. M., Ceccarelli, C., Castets, A., 
Wakelam, V., Caux, E., Parise, B., Teyssier, D. (2003). ApJ, 593, L51

\bibitem{} Chang, Q., Cuppen, H. M., Herbst, E. (2005). A\&A, in press

\bibitem{} Charnley, S. B.,  Tielens, A. G. G. M., Millar, T. J.
(1992). ApJ, 399, L71


\bibitem{} Chiar, J.E., Adamson, A.J., Kerr, T.H., Whittet, D.C.B.
(1995). ApJ, 455, 234

\bibitem{} Crapsi, A., Caselli, P., Walmsley, C. M.,
    Myers, P. C., Tafalla, M., Lee, C. W.,
    Bourke, T. L.  (2005). ApJ, 619, 379

\bibitem{} Dalgarno, A., Lepp, S.  (1984). ApJ, 287, L47

\bibitem{} Doty, S. D., van Dishoeck, E. F., van der Tak, F. F. S., 
Boonman, A. M. S. (2002). A\&A, 389, 446

\bibitem{} Draine, B. T. (1980). ApJ, 241, 1021

\bibitem{} Draine, B. T., Roberge, W. G., Dalgarno, A. (1983). 
ApJ, 264, 485

\bibitem{} Duley, W. W., Williams, D. A. (1984). 
\textit{Interstellar chemistry},  London, England and 
 Orlando, FL, Academic Press

\bibitem{} Dutrey, A., Guilloteau, S., Guelin, M.  (1997). A\&A, 
317, L55

\bibitem{} Ehrenfreund, P.,  Charnley, S. B. 2000, ARA\&A, 38, 427

\bibitem{} Evans, N. J., II  (1999). ARA\&A, 37, 311

\bibitem{} Evans, N. J., II, Rawlings, J. M. C., Shirley, Y. L., Mundy, 
L. G. (2001). ApJ, 557, 193

\bibitem{} Flower, D. R., Pineau des For\^ets, G. (1995). MNRAS, 275, 1049



\bibitem{} Fuente, A., Rizzo, J. R., Caselli, P., Bachiller, R., Henkel, C.
(2005). A\&A, in press (astro--ph/0411602)

\bibitem{} Galli, D., Walmsley, C. M., Gon\c alves, J. (2002). A\&A,
394, 275

\bibitem{} Geppert, W. D., Thomas, R., Semaniak, J., et al. (2004). 
ApJ, 609, 459

\bibitem{} Gibb, E. L., Whittet, D. C. B., Boogert, A. C. A., 
Tielens, A. G. G. M. (2004). ApJ, 610, L113

\bibitem{} Glassgold, A. E., Najita, J., Igea, J.  (1997). ApJ, 480, 344

\bibitem{} Glassgold, A. E., Najita, J., Igea, J.  (2004). ApJ, 615, 972

\bibitem{} Goldsmith, P. F., Langer, W. D., Velusamy, T.  (1999). 
ApJ, 519, L173

\bibitem{} Gould, R. J., Salpeter, E. E. (1963). ApJ, 138, 393

\bibitem{} Graedel, T. E., Langer, W. D., Frerking, M. A. (1982). 
ApJS, 48, 321

\bibitem{} Habart, E., Boulanger, F., Verstraete, L., Walmsley, C. M., 
Pineau des For\^ets, G. (2004). A\&A, 414, 531

\bibitem{} Hartquist, T. W., Dalgarno, A.,
    Oppenheimer, M. (1980). ApJ, 236, 182

\bibitem{} Hartquist, T. W., Williams, D. A. (1995). \textit{The Chemically 
Controlled Cosmos, Astronomical 
Stars} (Cambridge University Press)

\bibitem{} Hartmann, L. (2003). ApJ, 585, 398


\bibitem{} Hasegawa, T. I., Herbst, E. (1993). MNRAS, 261, 83

\bibitem{} Hasegawa, T. I., Herbst, E., Leung, C. M.
 (1992). ApJS, 82, 167

\bibitem{} Herbst, E. (1990). Angew. Chem. Int. Ed. Engl., 29,  595

\bibitem{} Herbst, E. (2000). In: \textit{Astrochemistry: From Molecular 
Clouds to Planetary Systems}, ed 
by Y. C. Minh \& E. F. van Dishoeck, p.147

\bibitem{} Herbst, E., Klemperer, W. (1973). ApJ, 185, 505

\bibitem{} Herbst, E., Leung, C. M. (1989). ApJS, 69, 271

\bibitem{} Hollenbach, D., Salpeter, E. E. (1971). ApJ, 163, 155

\bibitem{} Horn, A., M\o llendal, H., Sekiguchi, O., Uggerud, E., 
Roberts, H., Herbst, E., Viggiano, A. A., Fridgen, T. D.
(2004). ApJ, 611, 605 

\bibitem{} Jones, A. P., Williams, D. A. (1985). MNRAS, 217, 413

\bibitem{} J\o rgensen, J. K., Hogerheijde, M. R.,
Blake, G. A.,  van Dishoeck, E. F.,
Mundy, L. G., Sch\"ier, F.  (2004). A\&A, 415, 1021 

\bibitem{} Jura, M. (1975). ApJ, 197, 575

\bibitem{} Kaufman, M. J., Neufeld, D. A. (1996). ApJ, 456, 250

\bibitem{} Kuan, Y.-J., Huang, H.-C.,  Charnley, S. B., et al. 
 (2004). ApJ, 616, L27

\bibitem{} Kurtz, S., Cesaroni, R., Churchwell, E., Hofner, P., Walmsley, C. M.
(2000). In: \textit{Protostars and Planets IV}, eds Mannings, V., Boss, A.P., 
Russell, S. S., p. 299 (University of  Arizona Press)

\bibitem{} Lee, C. W., Myers, P. C., Tafalla, M. (2001). ApJS, 136, 703

\bibitem{} L\'eger, A., Jura, M., Omont, A.  (1985). A\&A, 144, 147

\bibitem{} Leung, C. M., Herbst, E., Huebner, W. F.
(1984). ApJS, 56, 231

\bibitem{} Li, D., Goldsmith, P. F. (2005). ApJ, in press (astro-ph/0412427)

\bibitem{} Luca, A., Voulot, D., Gerlich, D. {\it WDS'02 Proceedings of 
Contributed Papers, Part II}, 2002, MATFYZPRESS, p. 294

\bibitem{} Markwick, A. J., Charnley, S. B. (2004). In: 
\textit{Astrobiology: Future Perspectives},  
by P.Ehrenfreund et al. Leiden Observatory, The Netherlands Astrophysics 
and Space Science Library, Vol. 305 (Kluwer Academic Publishers), p. 33

\bibitem{} Markwick, A.J., Ilgner, M., Millar, T.J., Henning, Th.
 (2002). A\&A, 385, 632

\bibitem{} McCall, B. J., Huneycutt, A. J., Saykally, R. J.,  et al.  
(2003). Nature, 422, 500


\bibitem{} McKee, C. F. (1989). ApJ, 345, 782

\bibitem{} Menten, K. M., Walmsley, C. M., Henkel, C., Wilson, T. L., 
Snyder, L. E., Hollis, J. M., Lovas, F. J.  (1986). A\&A, 169, 271

\bibitem{} Mehringer, D. M., Snyder, L. E.  (1996). ApJ, 471, 897

\bibitem{} Millar, T. J., MacDonald, G. H., Habing, R. J. 
 (1995). MNRAS, 273, 25

\bibitem{} Morton, D. C. (1975). ApJ, 197, 85

\bibitem{} Myers, P. C. (1999). In: \textit{The Origin of Stars and Planetary 
Systems}, ed by C. J. Lada and N. D. Kylafis, p.67 (Kluwer Academic 
Publishers)

\bibitem{} Neufeld, D. A., Dalgarno, A. (1989). ApJ, 340, 869

\bibitem{} Nomura, H., Millar, T. J.  (2004). A\&A, 414, 409

\bibitem{} Nummelin, A., Whittet, D. C. B., Gibb, E. L., Gerakines, P. A., 
Chiar, J. E. (2001). ApJ, 558, 185

\bibitem{} Ohashi, N., Lee, S. W., Wilner, D. J., Hayashi, M.
(1999). ApJ, 518, L41

\bibitem{} Olofsson, H. (1984). A\&A, 134, 36

\bibitem{} Ossenkopf, V., Henning, Th. (1994). A\&A, 291, 943

\bibitem{} Pagani, L., Bacmann, A., Motte, F., et al. (2004). 
A\&A, 417, 605

\bibitem{} Palla, F., Stahler, S. W. (2002). ApJ, 581, 1194

\bibitem{} Palumbo, M. E., Geballe, T. R., Tielens, A. G. G. M.
(1997). ApJ, 479, 839

\bibitem{} Parise, B., Castets, A., Herbst, E.,
      Caux, E., Ceccarelli, C., Mukhopadhyay, I.,
      Tielens, A. G. G. M. (2004). A\&A, 416, 159

\bibitem{} Pauls, A., Wilson, T. L., Bieging, J. H., Martin, R. N.
 (1983). A\&A, 124, 23

\bibitem{} Pineau des For\^ets, G., Roueff, E.,  Schilke, P.,
    Flower, D. R. (1993). MNRAS, 262, 915

\bibitem{} Pirronello, V., Liu, C., Roser, J. E., Vidali, G.
(1999). A\&A, 344, 681

\bibitem{} Prasad, S. S., Huntress, W. T., Jr. (1982). ApJ, 260, 590

\bibitem{} Qi, C., Kessler, J. E., Koerner, D. W., Sargent, A. I.,
 Blake, G. A.  (2003). ApJ, 597, 986

\bibitem{} Remijan, A., Shiao, Y.-S., Friedel, D. N., Meier, D. S., Snyder, L. E
 (2004). ApJ, 617, 384

\bibitem{} Roberts, H., Herbst, E.,  Millar, T. J.  (2003). ApJ, 591, L41

\bibitem{} Rodgers, S. D., Charnley, S. B.  (2001). ApJ, 546, 324

\bibitem{} Rodgers, S. D., Charnley, S. B. (2003). ApJ, 585, 355

\bibitem{} Ruffle, D. P., Hartquist, T. W., Caselli, P., Williams, D. A.
 (1999). MNRAS, 306, 691

\bibitem{} Ruffle, D. P., Herbst, E. (2001). MNRAS, 322, 770

\bibitem{} Schilke, P., Walmsley, C. M.,
Pineau des For\^ets, G., Flower, D. R. (1997). A\&A, 321, 293

\bibitem{} Schutte, W. A., Khanna, R. K. (2003). A\&A, 398, 1049

\bibitem{} Sellgren, K., Smith, R. G., Brooke, T. Y.
 (1994). ApJ, 433, 179

\bibitem{} Semenov, D., Wiebe, D., Henning, Th. (2004). A\&A, 417, 93

\bibitem{} Shinnaga, H., Ohashi, N., Lee, S.-W., 
Moriarty-Schieven, G. H.  (2004). ApJ, 601, 962

\bibitem{} Snow, T. P., Witt, A. N. (1996). ApJ, 468, L65

\bibitem{} Spitzer, L. (1978). \textit{Physical processes in the interstellar 
medium} (New York Wiley-Interscience)

\bibitem{} Stahler, S. W., Palla, F. 2004, \textit{The formation of stars}, 
New York, NY: Wiley

\bibitem{} Stantcheva, T., Herbst, E. (2004). A\&A, 423, 241


\bibitem{} Tafalla, M., Myers, P. C., Caselli, P.,
    Walmsley, C. M. (2004). A\&A, 416, 191

\bibitem{} Takahashi, J., Williams, D. A. (2000). MNRAS, 314, 273

\bibitem{} Teixeira, T. C., Emerson, J. P., Palumbo, M. E.  (1998). A\&A, 
330, 711

\bibitem{} Thi, W.-F., van Zadelhoff, G.-J., van Dishoeck, E. F.
 (2004). A\&A, 425, 955

\bibitem{} Tielens, A. G. G. M. (2005). \textit{The Physics and Chemistry of 
the Interstellar Medium}, in press

\bibitem{} Tielens, A. G. G. M., Allamandola, L. J. (1987). In: 
\textit{Interstellar processes}, p. 397 (Dordrecht, D. Reidel Publishing Co.)

\bibitem{} Tielens, A. G. G. M., Hagen, W. (1982). A\&A, 114, 245

\bibitem{} Turner, B. E.  (1990). ApJ, 362, L29

\bibitem{} van der Tak, F. F. S., Schilke, P.,
    M\"uller, H. S. P., Lis, D. C., Phillips, T. G.,
    Gerin, M., Roueff, E.  (2002). A\&A, 388, L53

\bibitem{} van der Tak, F. F. S., van Dishoeck, E. F. (2000). A\&A, 358, L79

\bibitem{} van Dishoeck, E. F. (2004). ARA\&A, 42, 119

\bibitem{} van Dishoeck, E. F., Blake, G. A. (1998). ARA\&A, 36, 317

\bibitem{} van Dishoeck, E. F., Blake, G. A., Draine, B. T., Lunine, J. I.
(1993). In: \textit{Protostars and planets III}, p. 163

\bibitem{} van Dishoeck, E. F., Thi, W.-F., van Zadelhoff, G.-J. 
 (2003). A\&A, 400, L1

\bibitem{} van Zadelhoff, G.-J., Aikawa, Y., Hogerheijde, M. R., 
van Dishoeck, E. F. (2003). A\&A, 397, 789

\bibitem{} Vastel, C., Phillips, T. G., Yoshida, H.  (2004). ApJ, 606, L127

\bibitem{} Visser, A. E., Richer, J. S., Chandler, C. J.
 (2002). AJ, 124, 2756

\bibitem{} Viti, S., Collings, M. P., Dever, J. W., McCoustra, M. R. S., 
 Williams, D. A.  (2004). MNRAS, 354, 1141

\bibitem{} Wakelam, V., Caselli, P., Ceccarelli, C., Herbst, E., Castets, A.
(2004). A\&A, 422, 159

\bibitem{} Walmsley, C. M., Flower, D. R.,
      Pineau des For\^ets, G. (2004). A\&A, 418, 1035

\bibitem{} Ward-Thompson, D., Motte, F., Andr\'e, P. (1999). MNRAS, 305, 143

\bibitem{} Watson, W. D. (1974). ApJ, 188, 35

\bibitem{} Weingartner, J. C., Draine, B. T. (2001). ApJ, 563, 842

\bibitem{} Willacy, K., Langer, W. D. (2000). ApJ, 544, 903


\bibitem{} Willacy, K., Millar, T. J. (1998). MNRAS, 298, 562

\bibitem{} Williams, D. A., Williams, D. E., Clary, D., et al. (2000).
In: \textit{Molecular hydrogen in space}, ed by F.
Combes, and G. Pineau des For\^ets, p.99   
(Cambridge University Press)

\bibitem{} Williams, J. P., Bergin, E. A.,
    Caselli, P., Myers, P. C., Plume, R. (1998). ApJ, 503, 689


\bibitem{} Wyrowski, F., Hofner, P., Schilke, P., Walmsley, C. M., Wilner, D. J., 
 Wink, J. E.  (1997). A\&A, 320, L17

\bibitem{} Woon, D. E. (2002). ApJ, 569, 541

\bibitem{} Wright, M. C. H., Plambeck, R. L., Wilner, D. J. 
 (1996). ApJ, 469, 216


\end{thebibliography}
